\documentclass{cimento}

\setcopyright{J.~Beckers, F.~Kaspar and J.~Knollm\"uller on behalf of the COMPASS Collaboration}
\usepackage{graphicx}  

\title{Progress in the partial-wave analysis methods at COMPASS}
\author{J.~Beckers\from{ins:tum}\thanks{{julien.beckers@tum.de}},
F.~Kaspar\from{ins:tum}\from{ins:cl}\thanks{{florian.kaspar@tum.de}} \atque
J.~Knollm\"uller\from{ins:tum}\from{ins:cl}\thanks{{jakob.knollmueller@tum.de}}
for the COMPASS Collaboration}

\instlist{\inst{ins:tum}Technische Universit\"at M\"unchen, Physik Department, James-Franck-Straße 1, 85748 Garching bei M\"unchen
  \inst{ins:cl}Excellence Cluster Origins, Boltzmannstraße 2, 85748 Garching bei M\"unchen}

\begin{document}

\maketitle

\begin{abstract}
    We study the excitation spectrum of light and strange mesons in diffractive scattering. We identify different hadron resonances through partial wave analysis, which inherently relies on analysis models.
     Besides statistical uncertainties, the model dependence of the analysis introduces dominant systematic uncertainties.
     We discuss several of their sources for the $\pi^-\pi^-\pi^+$ and $K^0_S K^-$ final states and present methods to reduce them.
     We have developed a new approach exploiting a-priori knowledge of signal continuity over adjacent final-state-mass bins to stably fit a large pool of partial-waves to our data, allowing a clean identification of very small signals in our large data sets.
     For two-body final states of scalar particles, such as $K^0_S K^-$, mathematical ambiguities in the partial-wave decomposition lead to the same intensity distribution for different combinations of amplitude values.
     We will discuss these ambiguities and present solutions to resolve or at least reduce the number of possible solutions.
     Resolving these issues will allow for a complementary analysis of the $a_J$-like resonance sector in these two final states.
\end{abstract}

\section{Introduction}

At COMPASS, we study the spectrum of light mesons. They are produced as short-lived intermediate states in so-called diffractive-dissociation reactions, in which a high-energy hadron beam interacts strongly with a proton target. The decay daughters are measured in the spectrometer. In order to disentangle the contributions from the interfering states and extract their parameters, we perform a partial-wave analysis (PWA). In the following, we will present the advances in the PWA methods at COMPASS, in the $K_S^0K^-$ and $\pi^-\pi^-\pi^+$ final states. In this paper, we focus on the studies in the former channel, while a detailed report on the methods developed in the latter is given in \cite{meson_proceedings}.

\section{Partial-wave analysis method}

We will first summarise the method of partial-wave analysis, detailed in \cite{compass_review}. In order to gain information about the states produced in the diffractive-dissociation process, we model its intensity distribution in the invariant mass $m_X$ and in the $n$-body-phase-space variables $\tau_n$. We separate the amplitude describing the process into contributions with specific quantum numbers, so-called partial waves denoted by the index $a$, so that
\begin{equation}\label{eq:intensity_decomp}
    \mathcal{I}(m_X;\tau_n) = \left| \sum_{a} \, \mathcal{T}_a(m_X) \, \psi_a(m_X;\tau_n) \, \right|^2 \ .
\end{equation}
In eq.~\ref{eq:intensity_decomp}, each partial-wave amplitude is further subdivided into two parts. The decay amplitude $\psi_a$ describes the phase-space distribution of the daughter particles of a state with quantum numbers $a$. The production amplitude $\mathcal{T}_a$ is what we want to extract from the data, as it contains information about the produced mesonic states. To do this in a quasi-model-independent manner, the data is usually divided in kinematic bins of $m_X$, and then fitted with our model to extract the values of the production amplitudes in each bin. These may then serve as input to a second fit, in which their mass dependence is modelled in terms of resonances and the latter's parameters extracted.

\section{Ambiguities in the partial-wave analysis of the $K_S^0K^-$ final state}

\begin{figure}
        \centering
        \includegraphics[height=0.4\linewidth,clip]{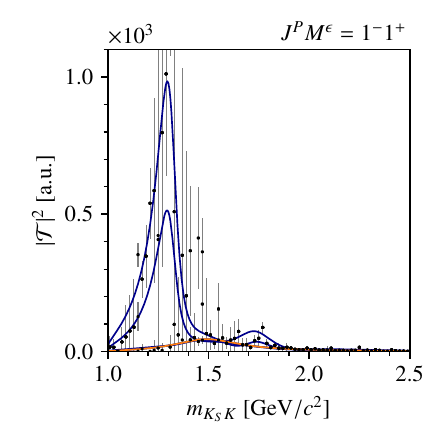}
        \includegraphics[height=0.4\linewidth,clip]{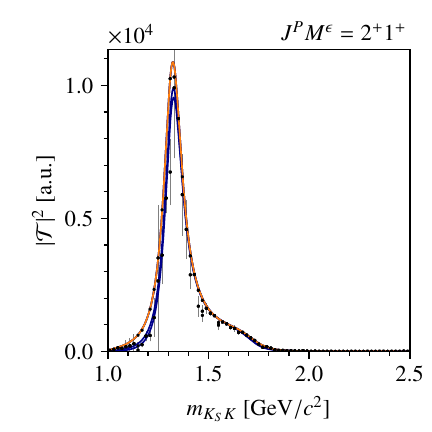}
        \includegraphics[height=0.4\linewidth,clip]{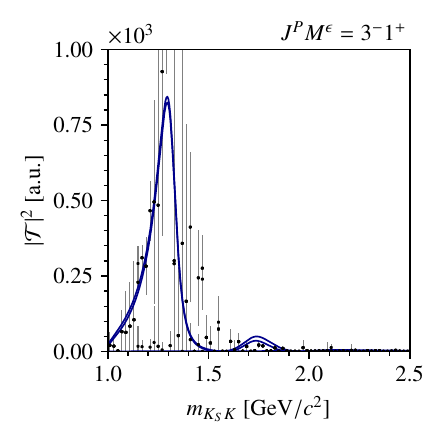}
        \includegraphics[height=0.4\linewidth,clip]{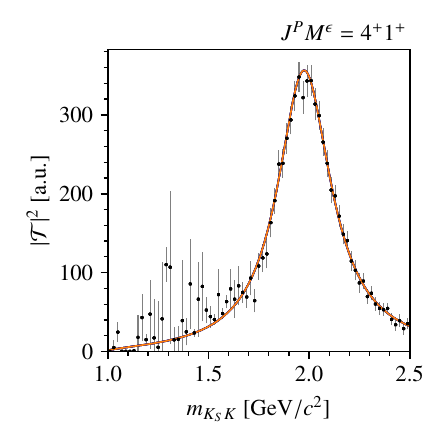}
        \caption{Partial-wave intensities of the ambiguous amplitudes. The model and calculated solutions are drawn as orange and blue curves respectively. The intensities obtained in the PWA fit are shown as black dots with grey error bars. See also \cite{meson_proceedings}.}
        \label{fig:amb_study}
    \end{figure}

\subsection{Non-uniqueness of the amplitude values}

For any final state with two spinless particles, such as $K_S^0K^-$, the decomposition of the intensity in eq.~\ref{eq:intensity_decomp} is not unique in each $m_X$ bin, and several sets of amplitudes $\{\mathcal{T}_{a}\}$ result in the same intensity distribution in phase space. To show this, and following \cite{chung_ambiguities}, we start by simplifying eq.~\ref{eq:intensity_decomp} to\footnote{We use the reflectivity basis for the amplitudes \cite{chung_spinDensityMatrix}. We neglect contributions with spin-projection $M \neq 1$ because of the dominance of Pomeron exchange at COMPASS energies \cite{compass_review} and the suppression of higher $M$ contributions. We drop indices $M^\varepsilon=1^+$.}
\begin{equation}\label{eq:amb_methods_intensity}
        \mathcal{I}(\theta,\phi) = \big| \sqrt{2} \, \underbrace{  \sum_{J=1}^{J_{\mathrm{max}}} \,\mathcal{T}_{J} \, Y_{J}^1(\theta,0) }_{a(\theta)} \, \big|^2 \ |{\sin{\phi}}|^2 \ .
\end{equation}
By expressing the spherical harmonics $ Y_{J}^1(\theta,0) $ as functions of the sole variable $u\equiv \tan{(\theta/2)}$, $a(\theta)$ in eq.~\ref{eq:amb_methods_intensity} can be written as a polynomial in $u^2$, which complex coefficients are linear combinations of $ \{ \mathcal{T}_J \} $. Finding the polynomial's $J_{\mathrm{max}}-1$ roots $r_k$, also called Barrelet Zeros \cite{barrelet}, enables us to write $a(\theta)$ as
\begin{equation}\label{eq:amb_methods_aplus_roots}
    (1+u^2)^{J_{\mathrm{max}}} \, a(u) = u \ c \prod_{k=1}^{J_{\mathrm{max}}-1} (u^2-r_k)
\end{equation}
by root decomposition. 
The product above enters only as absolute value in the intensity model from eq.~\ref{eq:amb_methods_intensity}. Therefore, the latter is invariant under complex conjugation of a single or several roots. Since the values of the roots $r_k$ depend non-linearly on $ \{ \mathcal{T}_J \} $, this operation evidently leads to different amplitude values. We can compute them from a starting set of amplitudes, by obtaining the roots of $a(\theta\,;\{ \mathcal{T}_J \})$ numerically. The new polynomials and sets of amplitude values can then be calculated from the $2^{J_{\mathrm{max}}-1}$ combinations of complex-conjugated roots.

\subsection{Pseudodata study of the ambiguities}

We have studied these ambiguities in the PWA of the $K_S^0K^-$ final state. To this end, we use a model containing four partial-wave amplitudes, parameterised close to physical amplitudes, i.e.~continuous in $m_X$ and containing resonant-like structures. The model amplitude of the $J^P=3^-$ wave is zero.
First, we compute the exact distributions of the ambiguous amplitudes, by sampling in $m_X$ and, at each point, using the values of the model amplitudes as input for the computation of the ambiguous solutions.
Fig.~\ref{fig:amb_study} shows the resulting intensity distributions. The model is drawn as an orange curve, and the ambiguous solutions in blue. Observing fig.~\ref{fig:amb_study}, we see that the ambiguous amplitudes are also continuous in $m_X$, but that the ambiguities lead to distributions that are significantly different from the starting model, especially in the smaller waves $J^P=1^-$ and $3^-$. We also note that the highest-spin wave (here $J^P=4^+$) is unaffected by the ambiguities. In a second step, we generate pseudodata according to the amplitude model and perform a PWA.\footnote{To ensure that we find all solutions, we perform a large number of fitting attempts with random starting values.} The resulting partial-wave intensities are shown as black dots in fig.~\ref{fig:amb_study}. Overall, the amplitudes found by the fit agree with the calculated distributions. Due to finite data however, the intensity distribution is distorted in some mass bins and the number of solutions is reduced. In addition, we see that the amplitude of the $J^P=4^+$ wave is still invariant.

\subsection{Resolving the ambiguities}

The ambiguities presented above are inherent to the $K_S^0K^-$ final state, but their influence on the partial-wave analysis can be alleviated in the COMPASS data by the choice of waveset.
Looking at the lower left plot of fig.~\ref{fig:amb_study}, we see that intensity appears in the ambiguous solutions of the $J^P=3^-$ wave although the model amplitude is zero. Leaving this wave out of the sum, effectively setting its amplitude to zero, hence does not prevent to correctly find the model solution. This, however, changes the intensity distribution of other solutions with a non-zero $3^-$ amplitude such that they are no longer indistinguishable.
This can be used in the COMPASS analysis: in our data, contributions with odd spins $J$ are suppressed. Excluding them from the waveset should matter only little in the PWA but effectively resolve the ambiguities in most mass bins.\footnote{We could also achieve this by manually sorting out "unphysical" solutions, i.e.~with significant intensity in these waves, but this way resolves the ambiguities directly in the fit procedure.} We have confirmed this by  performing a PWA fit of the pseudodata without the $3^-$ wave: there, only the model solution is found.

\section{Continuity for partial-wave analysis}
    We have seen, for the $K_S^0K^-$ final state, how ambiguities and noise can impact the quality of the fit results.
    This affects all such analyses, since the number of parameters in the fit models are large compared to the available data and the likelihood often shows multiple optima.

    We have developed a new approach that tries to reduce the uncertainties on the obtained solutions by making use of additional prior information.
    We know that the underlying signals are continuous and follow the reaction kinematics, meaning they should be suppressed both at threshold and for large final-state masses.
    In the binned approach, we extract the amplitudes in a quasi-model-independent way. We want our new model to keep this flexible behaviour.
    It is based on Information Field Theory (IFT) \cite{ift, corr_fields, nifty} and allows us to fit smooth curves to all bins simultaneously, thus enforcing our 
    prior requirements.
    We are also able to extend the model description with resonance parameterisations, allowing us to directly perform resonance fits in a single step.

\begin{figure}
        \centering
        \includegraphics[height=0.4\linewidth,clip]{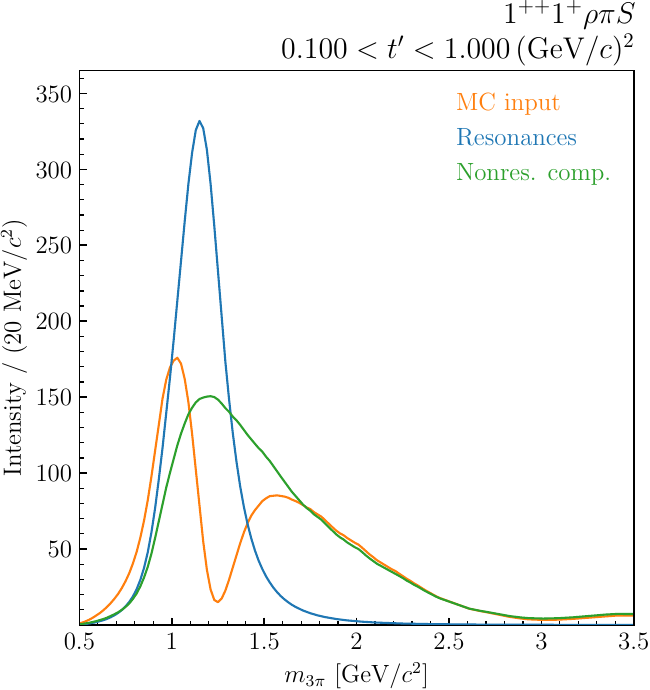}
        \hspace{0.5cm}
        \includegraphics[height=0.4\linewidth,clip]{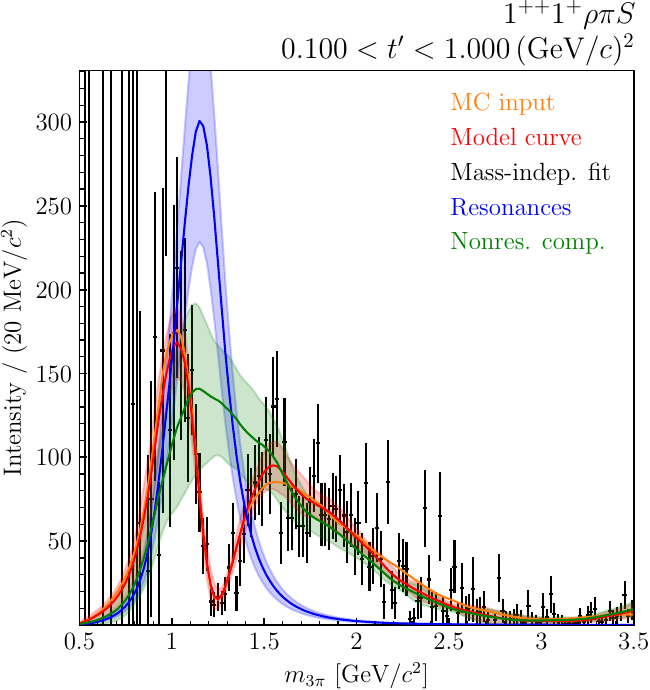}
        \caption{\emph{Left:} Intensity of input model (orange) split in resonant (blue) and non-resonant (green) components. \emph{Right:} Comparison of the two different fitting methods. The binned fit (black points) shows large statistical uncertainties. The new method (red) recovers the input model (orange) with high accuracy. It also correctly separates the components (blue and green). More examples can be found in \cite{meson_proceedings}.}
        \label{fig:nifty_fit}
    \end{figure}

    We studied the new method on simulated data of the $\pi^-\pi^-\pi^+$ final state.
    We generated data according to the continuous model and tried to recover the input using both the binned approach and the new method.
    Fig. \ref{fig:nifty_fit} demonstrates how combining information across multiple bins can drastically reduce the overall uncertainty of the fit as compared to the usual method.

\section{Outlook}

As discussed, the ambiguities in the partial-wave analysis of the $K_S^0K^-$ final state can be reduced. We are proceeding with the partial-wave analysis of the COMPASS $K_S^0K^-$ data and are extracting parameters of the $a_J$ states appearing therein. We are applying the continuous PWA method on the large COMPASS $\pi^-\pi^-\pi^+$ dataset to measure the contributing $a_J$ and $\pi_J$ resonances with increasingly higher precision and to improve our sensitivity to small signals. In addition, we are working on applying this novel method on other channels measured at COMPASS. Initial results on simulated data show that this approach may help to separate the different ambiguous solutions in $K_S^0K^-$.

\acknowledgments
 We would like to thank Philipp Frank and Torsten En{\ss}lin (both Max-Planck Institute for Astrophysics), as well as Stefan Wallner (Max-Planck Institute for Physics), who, together with the authors, made a first attempt to adapt NIFTy for partial-wave analyses. 
 We also thank the COMPASS Hadron Subgroup.\\
 Some of the results in this publication have been derived using the NIFTy package \cite{nifty}.\\
 Funded by the DFG under Germany's Excellence Strategy - EXC2094 - 390783311 and BMBF Verbundforschung 05P21WOCC1 COMPASS.

\end{document}